\def\etal{{\it et al. }}

 \documentclass[final,3p,times]{elsarticle}
\usepackage{lineno,hyperref}
\usepackage{amsmath}
\usepackage{color}
\usepackage{amssymb}
\usepackage{graphicx}
\usepackage{adjustbox}
\usepackage{lineno}

\usepackage{float}
\usepackage{lipsum}
\usepackage{mwe}

\usepackage{multirow}

\begin{document}
\begin{frontmatter}
	
	\title{Efficient machine-learning model for fast assessment of elastic properties of high-entropy alloys}
	
	\author[one]{Guillermo Vazquez\corref{mycorrespondingauthor}}
	\cortext[mycorrespondingauthor]{Corresponding authors}
	\author[one,two]{Prashant Singh\corref{mycorrespondingauthor}}
	\author[one]{Daniel Sauceda}
	\author[one]{Richard Couperthwaite}	
	\author[two]{Nicholas Britt}
	\author[three]{Khaled Youssef}
	\author[two,four]{Duane D. Johnson}
	\author[one,five,six]{Raymundo Arr\'{o}yave\corref{mycorrespondingauthor}}

	\address[one]{Department of Materials Science and Engineering, Texas A\&M University, College Station, TX 77843, USA}
	\address[two]{Ames Laboratory, U.S. Department of Energy, Iowa State University, Ames, Iowa 50011 USA}
	\address[three]{Materials Science \& Technology, Qatar University, Doha, Qatar}
	\address[four]{Department of Materials Science \& Engineering, Iowa State University, Ames, Iowa 50011 USA}
	\address[five]{Department of Mechanical Engineering, Texas A\&M University, College Station, TX 77843, USA}
	\address[six]{Department of Industrial and Systems Engineering, Texas A\&M University, College Station, TX 77843, USA}

	\begin{abstract}
		We combined descriptor-based analytical models for stiffness-matrix and elastic-moduli with mean-field methods to accelerate assessment of technologically useful properties of high-entropy alloys, such as strength and ductility. Model training for elastic properties uses Sure-Independence Screening (SIS) and Sparsifying Operator (SO) method yielding an optimal analytical model, constructed with meaningful atomic features to predict target properties. Computationally inexpensive analytical descriptors were trained using a database of the elastic properties determined from density functional theory for binary and ternary subsets of Nb-Mo-Ta-W-V refractory alloys. The optimal Elastic-SISSO models, extracted from an exponentially large feature space, give an extremely accurate prediction of  target properties, similar to or better than other models, with some verified from existing experiments. We also show that electronegativity variance and elastic-moduli can directly predict trends in ductility and yield strength of refractory HEAs, and reveals promising alloy concentration regions.
	\end{abstract}
	
	
\end{frontmatter}

\section*{Introduction}

{\par} High Entropy Alloys (HEAs) are novel class of alloys with intriguing electronic properties \cite{Zhang2020,Singh2018PRM} and often superior mechanical behavior \cite{Li2016, adma_201807142} attributed to near-equiatomic (5- 35 at.\%) mixing of multi-principle elements ~\cite{Yeh2004,Yeh2004a,Cantor2004}. The unique characteristic of single-phase solid-solution HEAs reveals a new and larger design space for complex solid-solution alloys \cite{ISHIZU2019102275,Zhang2020,George2019,IKEDA2019464}. Refractory-based HEAs (RHEAs), a special class of HEAs, show immense potential \cite{MIRACLE2017448} that brings these alloys under increased focus for their superior electronic \cite{Song2017,Singh2020} and mechanical properties (like high-strength \cite{ZHANG20141, MAITI201687}),  as well as very high melting temperature \cite{Senkov2012,Senkov2018,Singh2018,MIRACLE2017448}, fracture resistance \cite{Gludovatz1153},  lower kinetic rates and increased long-term high-temperature stability\cite{Chang2014}. RHEAs has expanded our search space needed for the discovery of viable candidates for a variety of current and anticipated applications requiring high ductility and fracture toughness, specific strength, and better mechanical performance at elevated temperatures.

{\par} Computational approaches  ({\it e.g.}, density-functional theory (DFT), molecular dynamics (MD) and/or CALPHAD [Calculation of Phase Diagrams]) or even empirical rules are used in combination to establish relationships between atomic ({\it e.g.}, electronegativity, atomic mass, atomic radii, etc.) and alloy features ({\it e.g.},  configurational entropy and mixing enthalpy) to design or tune the application relevant properties of RHEAs~\cite{PKL2008,Sheng2011,Steingrimsson2021,Pei2020}. However, excessive computational cost and uncertainty of first-principle methods limit the use of conventional approaches to explore exponentially large combinatorial design spaces (5$\times{10^{8}}$ compositions considering 25 different elements) \cite{Feng2017,LEE2021109260}. More recently, data-driven mehtods interfaced with machine-learning (ML) algorithms have enabled rapid filtering to reduce the vast alloying (Gibbs' compositional) space through fast-acting predictive models \cite{ML1,LEE2021109260,Zhou2019}. While successful, ML models oftentimes tend to be limited by the sparcity of the datasets available to train them~\cite{Zhang2018,deJong2016}. On other occasions, ML-based approaches have resulted in models that are difficult to interpret and are thus mostly 'black boxes' establishing connections between inputs and outputs through highly non-linear and convoluted mappings.

{\par} Here, we construct predictive models of the intrinsic mechanical behavior of RHEAs to determine critical trends that are key ML training information. Notably, if we consider the space of atomic features and an arbitrary set of operators used to integrate them, the dimensionality of such a problem can easily run into $10^8$-$10^9$ range. The discovery of the most informative feature subspace thus becomes an intractable problem using conventional dimensional-reduction approaches. Based on notions of Compressed Sensing (CS)~\cite{Ghiringhelli2017}, which solves for the sparse solution to an undetermined systems of equations, SISSO~\cite{Ouyang2018,Ouyang2019,Ouyang2017SISSO} uses Sure-Independent Screening \cite{Fan2008} (SIS), which selects at each iteration a new subset of features. Further reduction in the feature space and the development of a predictive model is carried out through Sparsifying Operators (SO), such as Least Absolute Shrinkage and Selection Operator (LASSO) \cite{Tibshirani1996} or $l_0$-norm regularized minimization in order to find the most sparse solution to a linear problem. The SISSO-based modeling is a form of symbolic regression, which uses trained analytical descriptors for property prediction. SISSO has already been shown to be a powerful automated-feature engineering (AFE) framework that enables the construction of accurate, interpretable predictive models for materials behavior, such as energy stability of inorganic solids \cite{Bartel2018, Bartel2019} as well as oxidation behavior in ceramics \cite{Sauceda2021}, among others. 

{\par} Knowledge of mechanical behavior is critical for the design of technologically useful alloys, however, the  main focus of ML models has been on thermodynamic stability with scant attention on developing fast and inexpensive models for mechanical properties \cite{Yin2019} due to unavailability of reliable databases. Elastic constants ({\it e.g.}, stiffness matrix (C$_{ij}$)) and other engineering quantities (such as yield-strength ($\sigma_{0y}$)) are extremely useful for materials design, yet, RHEA optimization has taken other design paths involving the use of simple (empirical) relations of atomic and thermodynamic properties \cite{Yang2012,Takeuchi,Guo2011a,Guo2011,Zhang2018}. Mechanical properties derived from elastic parameters can provide valuable insight into a RHEA's brittleness, stiffness, anisotropy, ductility, bonding character, and strength. However, our knowledge about features affecting important engineering quantities, such as strength, is very limited due to lack of efficient computational techniques. Therefore, methods or models capable of accurately predicting fundamental engineering quantities would greatly benefit the alloy design.

{\par} Here, we present descriptor-based analytical models, i.e., ``Elastic SISSO", for the fast exploration of mechanical properties over the vast  HEA space, which will also be useful as didactic tool for showcasing simple correlations between target properties---features identified from this method can in turn be used to train more complex 'black box' models, if so desired. We establish that dominant factors, such as high yield strength, derived from ML-predicted elastic parameters of body-centered cubic (bcc) RHEAs, consisting of 3$d$ (V), 4$d$ (Nb, Mo), and 5$d$ (Ta,W) transition-metal elements, can  be directly related to fundamental elemental quantities, {\it e.g.}, electronegativity variance ($\chi_{var}$), atomic-size differences ($\delta$), and formation enthalpy (E$_{form}$). Specifically, for Ta-W-Nb-Mo-V RHEA systems, we also discuss the relationship between phase stability, local environ correlation, and mechanical properties of alloying components. While there is not a consensus on a threshold that describes the ductility in HEAs using Pugh's ratio~\cite{Pugh1954}, Poisson's ratio, or Cauchy parameter~\cite{Pettifor2013}, to some extent all of these quantities relate to the ductility of the alloy and, with a clear trend established (and experimentally validated in subset of systems), they can be employed for design. We trained, tested, and developed descriptor-based analytical models using SISSO-based ML to predict mechanical properties of solid-solutions. This systematic study for various mechanical properties will benefit material science community in accelerating search for technological useful HEAs.

{\par}Different models and designs had been proposed to predict or induce ductility in RHEAs, for example, whether intrinsic or extrinsic, design approaches have been taken to overcome brittleness \cite{Wu2019}. Notably, the Valence Electron Concentration (VEC) has also been linked to ductility, for example, Qi \etal studied intrinsic ductility in BCC alloys by comparing shear instability and crack initiation using DFT  \cite{Qi2014} by optimizing average VEC \cite{Sheikh2016}. Chen \etal also provided distinctions between different ranges of VECs \cite{Chen2018a}. \newline

\section{Methods}

\subsection{DFT Elastic Database} 
The stiffness matrix of binary and ternary Nb-Mo-Ta-W-V based alloys were calculated using DFT-based stress-strain approach  \cite{LePage2002,Shang2007}, as implemented in the plane-wave pseudo-potential Vienna Ab-initio Simulation Package (VASP) \cite{Kresse1993,Kresse1999}. Both special quasirandom structure (SQS) \cite{Wei1990} and supercell random approximates (SCRAPs) \cite{Singh2021} methods were used to mimic homogeneously random disordered alloys. The Perdew, Burke, and Ernzerhof (PBE) generalized gradient approximation \cite{PerdewPBE1996} to DFT was used with a energy cut-off of 520 eV. Full (volume and ionic) optimization and charge self-consistency were done on (2 $\times$ 2 $\times$ 2)  and (4 $\times$ 4 $\times$ 4) Monkhorst-Pack \cite{Monkhorst1976} $k-$mesh for Brillouin zone integration, respectively. A high level of convergence criteria was set for elastic property calculation in both energy and forces, i.e., 10$^{-6}$ eV and 10$^{-6}$ eV/\AA. 

{\par} As supercells used have differing structure along x, y, and z axes, when elastic distortions are applied in DFT calculations the elastic constants C$_{11}$, C$_{22}$, and C$_{33}$, for example, will be slightly different, rather than exactly equal due to average cubic symmetry for a homogeneous solid solution. Hence, we average appropriately (see below). Here, we applied 12 independent deformations to each investigated structure, one each for the normal and shear strains and  two strains with opposing directions. Combination of binary and a ternary sets were chosen to map the stiffness data on the 5-dimensional HEA space. For binaries, ten subsystems were sampled using a 60-atom supercell in steps of 10\%  (with a single sub-step of 5\% at composition $(0.25,0.75)$), while ternaries were sampled using a 64-atom supercell with base composition of $(0.125,0.125,0.75)$ and $(0.25,0.375,0.375)$. The equivalent compositions were generated by shuffling the atom positions for both binary and ternaries making it 10 different cases per binary and ternary, which makes up to a total of 170 combinations (110 binary and 60 ternary). The macroscopic value of the elastic stiffness coefficients was then approximated by averaging the three independent elastic constants C$_{ii}$ and C$_{ij}$ (with i$\ne$j) in a cubic crystalline system \cite{Tian2015,Tasnadi2012}, i.e., C$_{11}^{avg}$=(C$_{11}$+C$_{22}$+C$_{33}$)/3, C$_{44}^{avg}$=(C$_{44}$+C$_{55}$+C$_{66}$)/3, and C$_{12}^{avg}$=(C$_{12}$+C$_{13}$+C$_{23}$)/3. The bulk (K)-, shear (G)-, and Young's (E)-moduli, Poisson's ratio ($\nu$) were approximated using the Voight-Reuss-Hill approximation\cite{Hill1952} as detailed by Wu \etal for the cubic system \cite{Wu2007}.

\subsection{Phase stability analysis.}
Low-fidelity high-throughput formation enthalpy calculation for Nb-Ta-Mo-W-V HEAs were performed using DFT-based Green's function Korringa-Kohn-Rostoker electronic-structure methods \cite{DDJ1,DDJ2}, in which the coherent-potential approximation accounts properly for averaging over all chemically disorder configurations in structures \cite{DDJ3}. The gradient-corrected Perdew, Burke, and Ernzerhof (PBE)  exchange-correlation functional \cite{PerdewPBE1996} was used within each atomic site for charge distributions and the total energy. A semi-circular contour and Gauss-Laguerre quadratures using 24 complex energies was used for integration. A 24$\times$ 24 $\times$ 24 Monkhort-Pack\cite{Monkhorst1976} $k$-mesh was used for Brillouin-zone integration of bcc Nb-Ta-Mo-W-V HEAs.

\subsection {Misfit volume calculation for strength prediction.} 

The misfit volume of the alloys can be calculated as the derivatives of atomic volume of elements in HEAs with respect to composition as $\Delta V_n = \frac{{\partial V_{{\mathrm{HEA}}}}}{{\partial c_n}} - \mathop {\sum}\limits_{m = 1}^N {c_m} \frac{{\partial V_{{\mathrm{HEA}}}}}{{\partial c_m}}$. The expression can be derived from the expression $\sum_{n}c_{n}=1$ using conditions $V_{{\mathrm{HEA}}} = V_{{\mathrm{HEA}}}(c_1,c_2,...,c_{N - 1})$ and $\partial V_{{\mathrm{HEA}}}{\mathrm{/}}\partial c_{N} = 0$. \cite{maresca2020mechanistic}

\section{Results}

\subsection{Elastic SISSO - Featurization.}

\begin{figure}
    \centering
    \includegraphics[width=0.75\columnwidth]{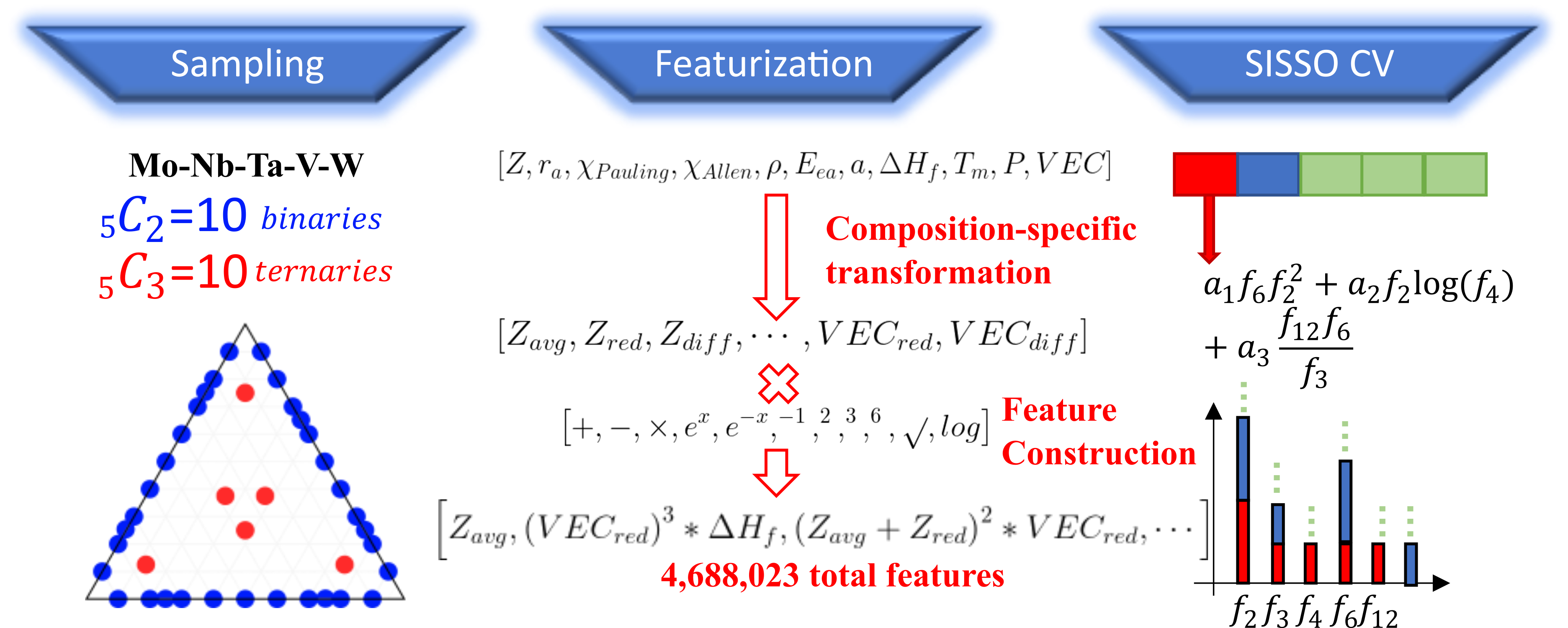}
    \caption{Schematic diagram for sampling, feature construction, and feature analysis processes.}
\end{figure}

Featurization of the alloy sampling is carried out via a stoichiometric approach, extrapolating elemental properties. Bartel \etal proposed these alloy-specific transformations in their Gibb's energy model whereby three different features are retrieved for each elemental property, using the stoichiometrically-weighted average (avg), the stoichiometrically-weighted harmonic mean (red), and the stoichiometrically mean difference (diff) of the atomic feature in question \cite{Bartel2018}, which are showcased in supplemental Table~3. 

We use atomic properties: atomic number, period, atomic radius, Pauling (molecular) and Allen (solid-state) electronegativity, density, heat of formation, BCC lattice constant, melting point, electron affinity and valence electron count $({\it i.e.}, Z, P, r_a, \chi_{Pauling},\chi_{Allen}, \rho, \Delta H, a, T_m, E_{ea},$ and $VEC)$. These 11 elemental features -- once transformed -- make up the 33-dimensional primary alloy feature space fed into the SISSO AFE framework. The feature construction was carried out by applying the operator set $\left [ +, - , \times,e^x,e^{-x},^{-1},^2,^3,^6,\sqrt{ },log \right ]$ to the 33 alloy-specific features. As shown in schematic Fig.~{1}, all features were created by combination of the primary features and operators, which were recursively added to the feature set two times.

\subsection{Training, model-generation, and cross-validation.}
During each SISSO iteration, the SIS method selects 1,000 features, which are added to a subset to which the $l_0$ norm regularized minimization Sparsifying Operator is applied so as to obtain the best linear regressor with a number of terms equal to the current iteration out of this subset by doing this every iteration ($1D$, $2D$, and $3D$ descriptors). Consequently, for the last descriptors $\left ( {3000 \atop 3} \right ) \sim 4.5 \times 10^9$ combinations were tested. Out of the 3 descriptors (1D, 2D, and 3D), the 3D descriptor shows the best accuracy, both in training and cross-validation (CV)  estimation, which indicates that over-fitting, a recurring caveat in sparse regressor, may start occurring at dimensions higher than 3. 


{\par} The SISSO models used in this work were put to test in a couple of different ways: (i) a 5-fold/10-fold CV to understand the over-fitting and frequency analysis of features found in CV, and (ii) the descriptors trained on full data. Thus, we only discuss the most complete descriptors and apply them here for further sampling of mechanical properties over complex composition space. As the systems have cubic symmetry, the SISSO model were retrained for two independent elastic constants [C$_{11}$; C$_{12}$] along with (K; G; E; $\nu$; (inverse) Pugh's ratio, G/K; C$_{44}$) that can be derived from Voigt-Reuss-Hill approximations in terms of C$_{11}$ and C$_{12}$.\cite{Hill1952}


\subsection{Elastic parameter C$_{11}$ descriptor $-$ Cross validation and error analysis.}
\begin{gather}
\begin{aligned}\label{Eq_C11}
\scalebox{0.9}{$C_{11}=  11.6 + 1.14 \left[{VEC_{avg}^{3}}\sqrt{ln(P_{avg})}\right] + 47.7 \left[VEC_{diff}^{36} \right] - 0.029 a_{avg}^6 \left[ VEC_{red} +VEC_{diff}\right]$}
\end{aligned}
\end{gather} 

  \begin{figure}
    \centering
    \includegraphics[width=1\columnwidth]{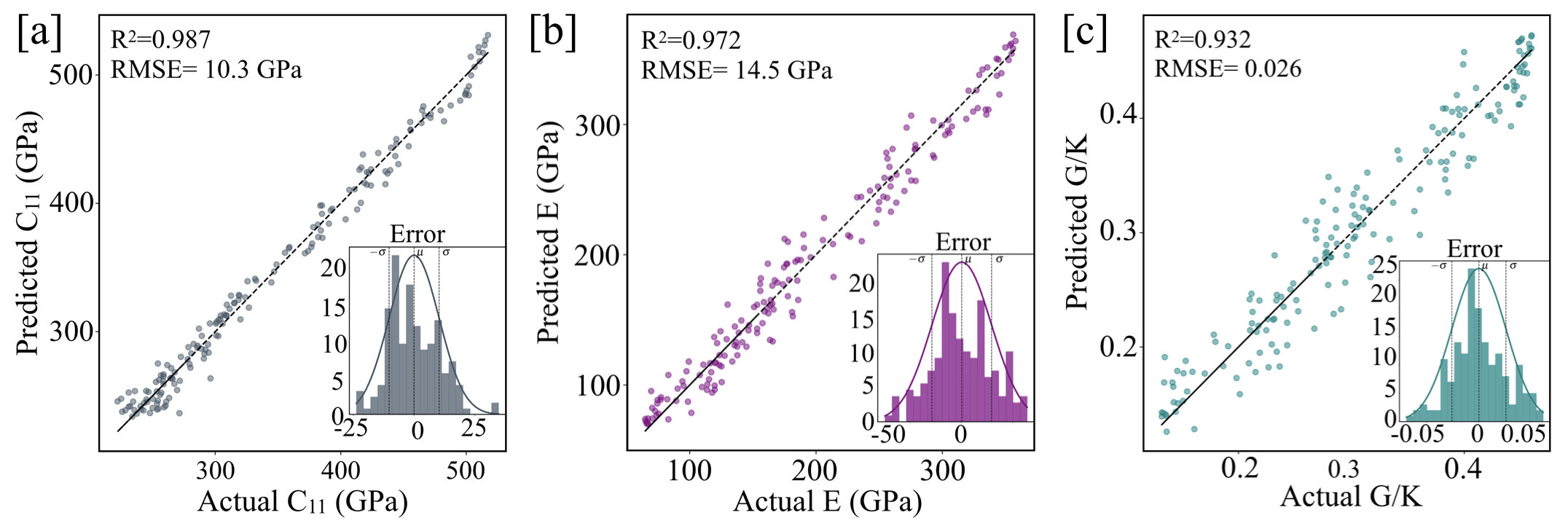}
    \caption{Actual (DFT) vs predicted (ML) (a) stiffness parameter (C$_{11}$; GPa), (b) Young's moduli (E; GPa), and (c) (inverse) Pugh's ratio (PR=G/K) from five-fold cross-validation test. Inset in a-c shows the error distribution around the mean, which is virtually zero for all the three cases.}
\end{figure}

{\par}The parity plot in Fig.~{1}a shows a good predictability even for a small number of folds. The 5-fold CV results for the stiffness matrix parameter C$_{11}$, for which the coefficient of determination indicates that an independent SISSO descriptor trained in 80\% of the data is able to determine the rest 20\% consistently. The inset graph shows the error distribution for the model and an approximation for the probability density, which follows a normal distribution. While for the evaluation of these results five different analytical models were used, Eq.~\ref{Eq_C11} shows a final model using all data available for the extrapolation to a higher dimension space in the next section.

\subsection{Young's moduli (E) descriptor $-$ Cross validation and error analysis.}
\begin{gather}
\begin{aligned}\label{eq:young}
\scalebox{0.9}{$ E   =  -469+1.31\frac{VEC_{avg} T_{m,avg}}{\sqrt{\Delta H_{f,avg}}} -0.335\frac{VEC_{red} \Delta H_{f,avg}}{VEC_{avg}-VEC_{diff}}+1.37\frac{VEC_{diff}}{E_{ea,diff}\chi_{Pauling,diff}^2}  $}
\end{aligned}
\end{gather}

{\par}Figure~{2}b corresponds to the results for the Young's moduli ($E$), the coefficients of determination for E is overall lower than for C$_{11}$. In the case shown here, the 5-fold CV, the error distribution once again has nearly a zero mean error. The final model is shown in Eq.~\ref{eq:young}, which in comparison appears to be more connected to the composition-specific features of the electronegativities.

\subsection{Pugh's ratio (PR) descriptor$-$ Cross validation and error analysis.}
Lastly, the inverse Pugh's ratio model in  Fig.~{2}c is performing with less accuracy than the C$_{11}$ and E. Evaluating a single model rather than two adds uncertainty instead of performing at the lowest accuracy of the two models $G$ and $K$ (see Table.~S1). The final analytical equation for  Pugh's ratio is shown in Eq.~\ref{Eq_GKratio}.
\begin{gather}
\begin{aligned}\label{Eq_GKratio}
\scalebox{0.9}{$ PR =  3.82 -0.293  \left[\frac{\chi_{Allen,avg} a_{avg}}{\sqrt{VEC_{avg}}}\right]
-0.0191\left[ \frac{exp\left(VEC_{diff}\right)}{\chi_{Pauling,avg} E_{ea,red}} \right]
-0.105\left[ \frac{\rho_{red}}{E_{ea,diff} \chi_{Pauling,diff} \Delta H_{f,avg}} \right]$}
\end{aligned}
\end{gather}

{\par}While Bartel \etal trained a SISSO descriptor for the thermodynamic properties of crystalline compounds by supplying compound-specific properties, the final descriptor for Gibbs' energy only chose the reduced mass from these transformed features, besides calculated atomic volume and temperature~\cite{Bartel2018}. Descriptors shown here are dependent solely on compound-specific features, we hypothesize the complexity of binaries and ternaries compounds elastic properties can be captured by non-linear features arranged in a linear combination. The complexity of the models is then assumed to translate to higher dimensions. Elastic stiffness constants naturally arise from tensor analysis of the strain-stress response of materials and the factors governing this response get as complex as the number of constituents grows. The SISSO-chosen descriptors shown here deviate enough from the linearity of the common rule-of-mixtures estimation to deliver a better approximation as the complexity of the alloy grows. The compromise for accuracy when using this descriptor is then compared to results of  DFT.

\subsection{Target property descriptor analysis.}
If a feature is selected in each term of five three-dimensional (3D) SISSO equations in 5-fold cross-validation test, then the highest frequency a feature can get is 15. To better understand these results, we present a descriptor analysis in Fig.~{3}, which consists of counting features by appearance inside a term for every equation and displaying 10 features with highest frequency. The most frequent features are the one those consistently appear in the analytical 3D models. The final equations may resemble the alloy-specific features chosen by the CV method, however, this selection of features is not guaranteed in other cases.

{\par}Final models (Eq. 1-3) and CV analysis within SISSO framework show some common alloy-specific features. For example, majority of descriptors include the VEC and Pauling electronegativity as a shared feature. The feature analysis in Fig.~3 is provided to understand feature presence and different mathematical operations in SISSO models. Moreover, the ranking of the atomic properties by appearance in the cross-validation method is not analogous to feature importance since they are not the operator coupled features used in SISSO, but it gives a qualitative idea about the basic building blocks of the descriptors, i.e., the composition-weighted features.

While it appears that the accuracy is correlated to the confidence of the CV in selecting the same alloy-specific features as in analytical SISSO models, this is not necessarily true for each case. For example, C$_{11}$ and E have higher confidence in their top features in contrast to model for the Pugh's ratio.  Yet, the diverse feature selected in each run are inherently correlated that create multiple top performing descriptors once coupled with the sparsyfying operators in SISSO method. Fig.~{3} reveals the elemental features such as VEC that appears in each model, which was found most significant for predicting elastic properties and consistently appears in the cross-validation.

 \begin{figure}
    \centering
    \includegraphics[width=0.95\columnwidth]{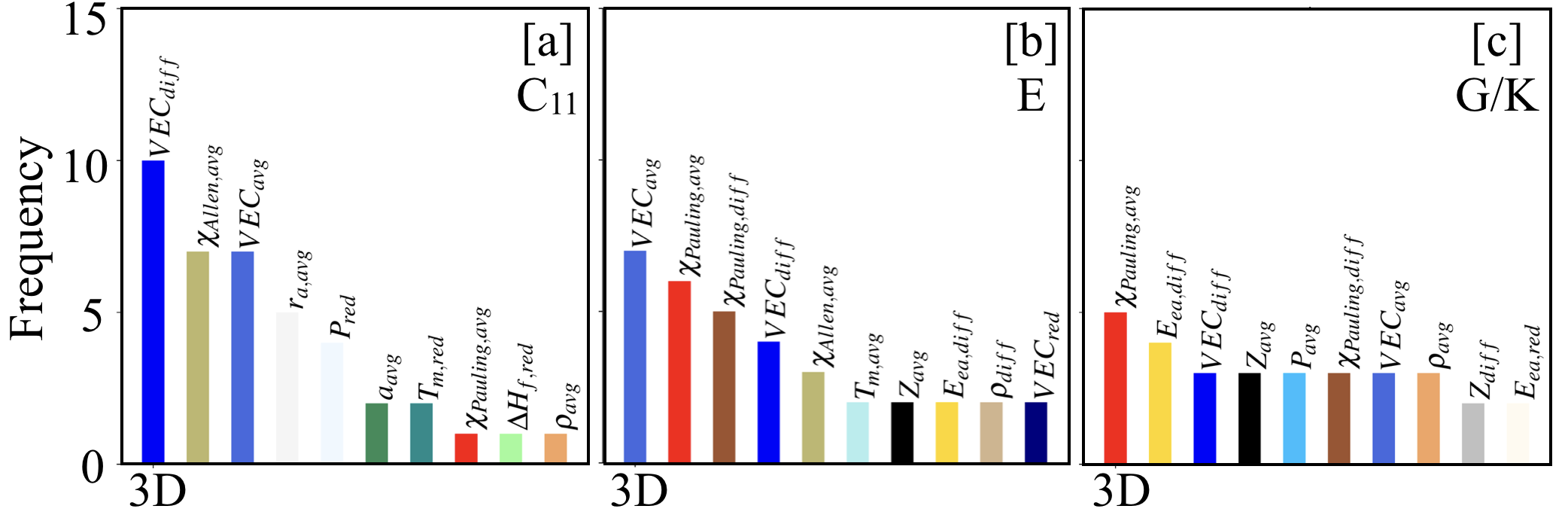}
    \caption{(a-c) Descriptor analysis ranks feature appearance in C$_{11}$, E, and Pugh's ratio in the CV method}
\end{figure}

Analytical models such as those available in Eq.~[1-3] and Table.~S1 are easy to use, which will expand new non-data-intensive approaches for rapid space exploration of complex materials including high-entropy alloys. For engineering purposes where the ductility is an important performance factor, an inexpensive analytical model can be useful for composition optimization from large alloy design space.

\section{Application to the Quinary Alloy}
\subsection{Interpretation of target properties and its interdependence on physical features.}

Unlike other 'black-box' ML methods, it is possible to create a direct relation with a specific feature from the get-go by analyzing the analytical function. We can go one step forward and evaluate a quinary random sampling and reveal how a subset of features (for each model, e.g., C$_{11}$, and E)  contributes to the ML models given in Fig.~{4}. These analytical models for elastic parameters were evaluated over 100,000 points in the Mo-Nb-Ta-V-W RHEA space.

{\it Stiffness constant} {C$_{11}$}:~Analytical model for the C$_{11}$ in Eq.~{1} strongly depends on the VEC, as obvious from Fig.~{4} (top-panel). The symmetry of the system constrains the values of all the weighted transformations for VEC to be completely correlated, i.e., the relation of C$_{11}$ with the average, harmonic average and difference-average is always the same. Therefore, the average of the periods (in periodic table) are correlated to some extent with the number of electrons, especially, when only looking at five different elements distributed over 3 different periods. This feature is present in the final descriptor, but only harmonic average of periods appeared during  cross-validation. The period in this system is constrained to integer values, which indicates an even stronger correlation between the different methods of averaging. The second term is filled with average difference of features, the valence electron, melting point and density. Here, the average difference between lowest and highest points indicate a higher C$_{11}$, so not only high differences between constituents yields a high C$_{11}$, but also small difference stabilizes the high stiffness. For example, a maximum in C$_{11}$ was obtained for the optimal range of average lattice parameter (via Vegard's rule), i.e., 3.15\AA~and 3.20\AA~. In supplementary Fig. 3, we  provide a short discussion on the usefulness of descriptor-based analytical approaches for property prediction in complex alloys over that of simple rules. The observations that the difference on electronegativity correlates well to mechanical properties of HEAs in this work agrees well with existing literature.\cite{oh2019engineering}

\begin{figure}
    \centering
    \includegraphics[width=0.9\columnwidth]{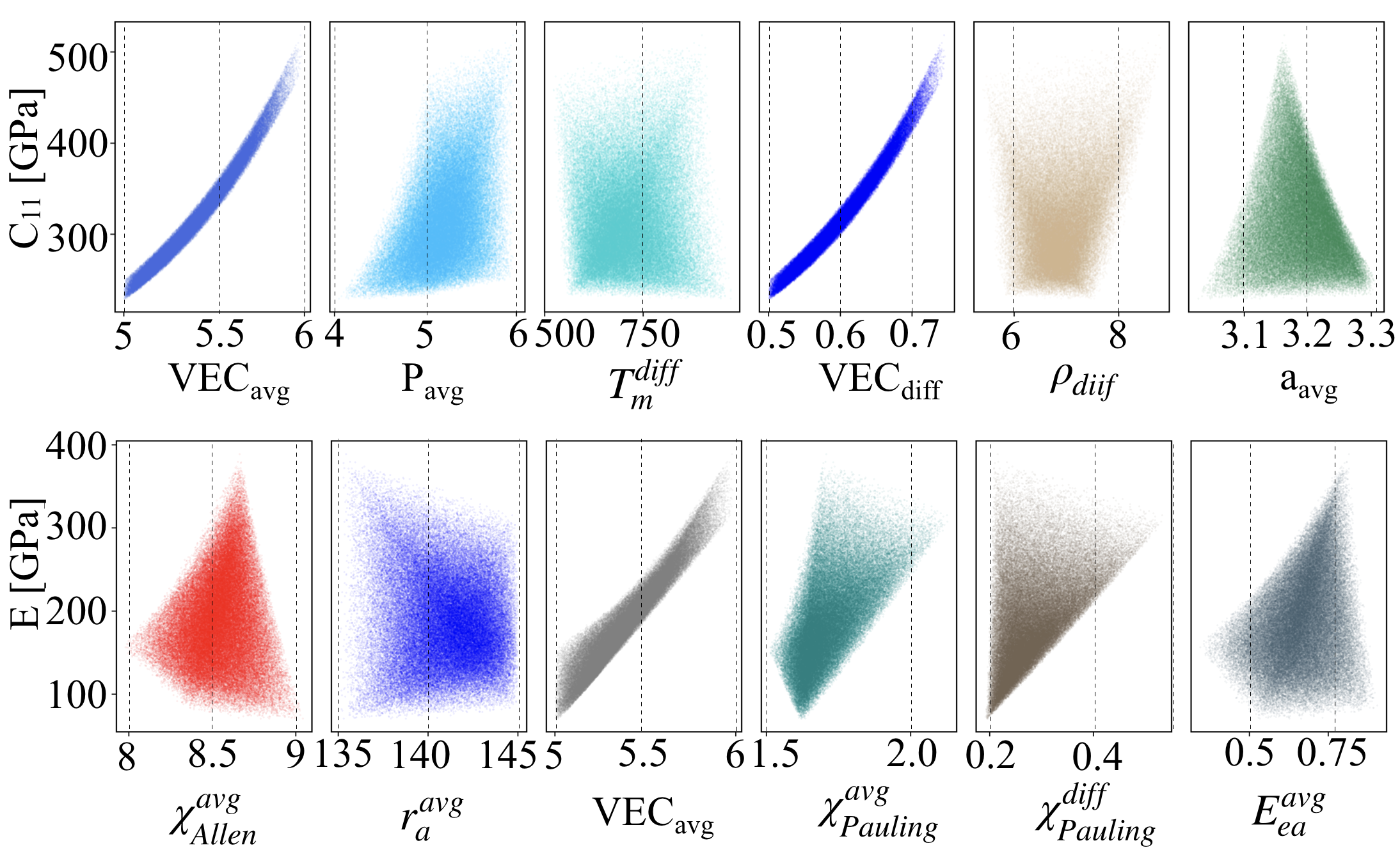}
    \caption{Chosen features (top-panel) stiffness matrix element (C$_{11}$), and (bottom-panel) Young's moduli ($E$) descriptor.}
\end{figure}

{\it Young's modulus E:}~Descriptor for $E$ in Eq.~{2} strongly depends on  VEC and $\chi$, as shown in Fig.~{4} (bottom-panel). As expected, average electronegativities have different correlation to $E$ due to the difference in their physical origin. Allen average electronegativity has an optimum value for  $E$ that is largest, while Pauling has an overall positive correlation. The $\chi_{Pauling}^{diff}$ shows good correlation with $\chi_{Pauling}^{avg}$ [see  Fig.~{4} (top-panel)] except for the onset where $\chi_{Pauling}^{diff}$ converges at single point while $\chi_{Pauling}^{avg}$ shows wide distribution. Moreover, we need to be cautious here because former represents the averages of absolute $\chi$ and later is average of difference in $\chi$. On the other hand, the electron affinity shows a higher range of stiffness at elevated values, and VEC shows a similar correlation with C$_{11}$. This last statement is true for all features, and $E$ and C$_{11}$ are so closely correlated that the feature-target correlations shape are  similar.

{\par}The average of lattice constant has proven to deviate by a small margin from the experimental lattice parameter, which follows a negative slope after a maximum value of 3.17\AA~for both C$_{11}$ and $E$. Gorban \etal \cite{krapivka2017high} also observed the same trend, which was attributed to size mismatch and weakening atomic interactions at a larger lattice parameter. A positive trend in modulus of elasticity at a higher electron count has been observed, {\it i.e.}, 6e$^{-}$ \cite{gorban2018role,krapivka2017high}, SISSO models presented in this work show similar behavior.

\subsection{Pugh's ratio -- evaluating derived parameters from stiffness matrix.}

{\par}  The inverse Pugh's ratio (G/K) and Cauchy pressure (C$_{12} -$ C$_{44}$) are empirically defined quantities, which are commonly used  to predict the ductility of complex alloys. Originally,  Pugh \cite{Pugh1954} discussed the ratio of (G/K $<$ 0.57 or K/G $>$1.75) and found it correlated with the ductility of elemental metals (fcc, bcc, and hcp), where bulk moduli (K) assesses the resistance to fracture while shear modulus (G) the tendency for increased fracture resistance after the onset of plastic deformation. Indeed, Gschneidner Jr. \emph{et al.} used existing experimental data from large number of materials to categorize brittle-to-ductile transition at $\text{G/K}<0.57$,\cite{gschneidner2003family} where the higher (lower) ratio indicates ductility (brittleness). Here, we use the inverse Pugh's ratio (PR=G/K) simply to span a $0 \le \text{PR} \le 1$ range, and ductility is indicated by $\text{G/K} < 0.57$.
For HEAs, the randomly sampled 100,000 compositions in the quinary alloy space (composition vector) were mapped on a two-dimensional space using a t-distributed Stochastic Neighbor Embedding (t-SNE) scheme in Fig.~5a, where the clustering algorithm separates the unary-rich composition. The percentage of composition is more evenly distributed away from t-SNE corners in Fig.~5a.  The (inverse) Pugh's ratio (G/K) in Fig. 5b is below the anisotropy-specific ductile-to-brittle critical point, as evaluated using the descriptor in Eq. (3). The HEAs have a positive Cauchy pressure ($\text{C}_{12} - \text{C}_{44} < 0$), as shown in Fig. 5c. 

{\par} Recently, Senkov et al. \cite{Senkov123} expanded the correlation between these two quantities by taking the ratio of Cauchy pressure and bulk moduli. The anisotropy-specific critical value for the Pugh's ratio in Fig.~5b was found in good agreement with Pettifor's condition for incipient brittleness, i.e., $\text{C}_{12} - \text{C}_{44} < 0$.  Although this would be considered ductile for elements and simple compounds, the ductility of HEAs is more complex, and such predictions must be seen as qualitative. Physically, the Pugh's ratio predicted for RHEAs in Fig.~5b shows the ability of an alloy to deform, where smaller value ($\text{G/K}<0.57$) shows higher deformability (ductile) while higher value ($\text{G/K}>0.57$) signifies lower deformability (brittle). Models for Pugh's ratio and Cauchy pressure show similar trends, i.e., the minimum for one in the quinary space is the maximum for the other. Notably, both models predict high ductility for HEAs in the same composition region. The model also shows a soft gradient between consecutive alloys in the t-SNE. Similarly, Pugh's ratio follows expected positive correlation with $VEC$, meaning that the alloy system is expected to show higher ductility as the $VEC_{avg}$ decreases. Both the bulk moduli (K) and shear moduli (G) have a positive slope with $VEC_{avg}$, whereas $G$ has a steeper increment. This suggests that  the alloy tends to get harder for compression or shear. Yet, the shear mechanism will be more affected as the bond energy of the crystal becomes stronger. The average and the difference of the lattice parameter show different correlations to the Pugh's ratio.

\begin{figure}
    \centering
    \includegraphics[width=0.8\columnwidth]{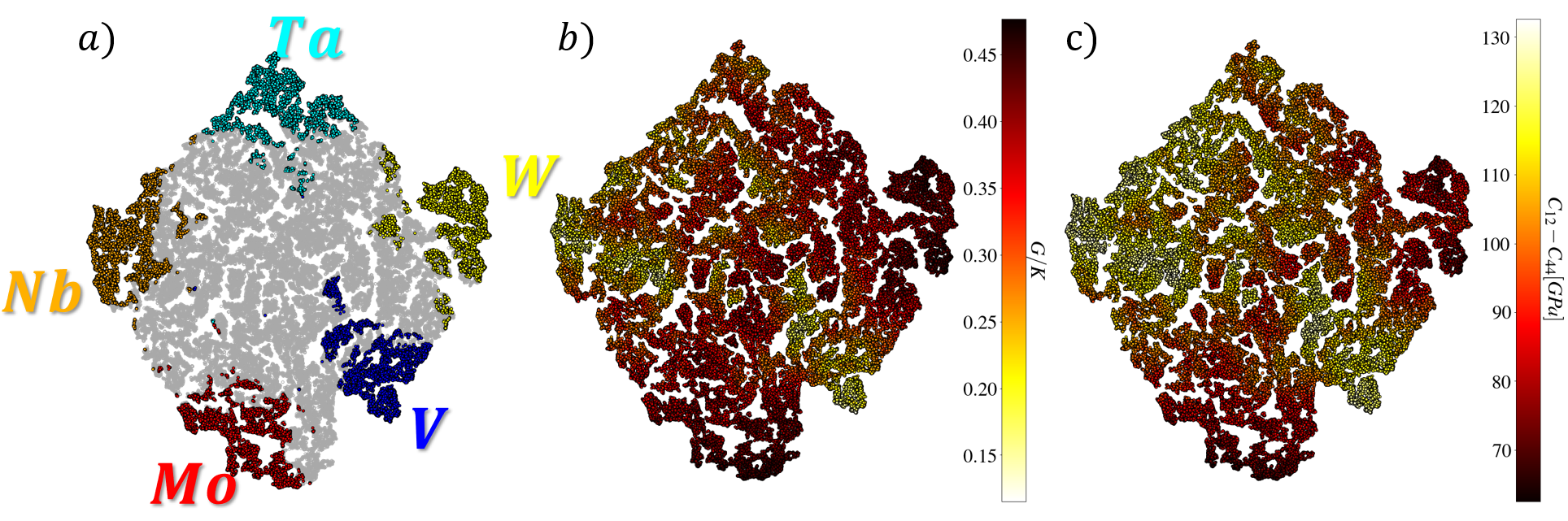}
    \caption{(a) t-SNE dimensional reduction of the quinary space with highlighted unary-enriched zones, the color scheme indicates that this point contains more than 50\% molar of this respective element, (b) Pugh's ratio, and (c) Cauchy pressure C$_{12} - $C$_{44}$ mapped in top of t-SNE, as predicted by the SISSO models.}  
\label{fig:sisso_1}
\end{figure}

\subsection{SISSO prediction of C$_{44}$ and isotropy dilemma.}
DFT methods often do not provide accurate estimate of shear constant (C$_{44}$) in isotropic disordered solids such bcc and fcc alloys. However, the C$_{44}$ can be directly estimated using the independent stiffness constants (C$_{11}$ and C$_{12}$) in a system with cubic symmetry, assuming isotropic behavior. To exemplify this, we show in Fig.~6  a comparison between the ML predicted C$_{44}$ and others derived from relation [$\text{C}_{11}-\text{C}_{12}]/2$. Notably, the C$_{44}$ derived from the isotropic assumption correlates well to ML predictions. The small prediction error of 7.34 GPa (RMSE) in C$_{44}$ corresponds to an average deviation from the model of 8.6\%.

\begin{figure}
    \centering
    \includegraphics[width=0.35\columnwidth]{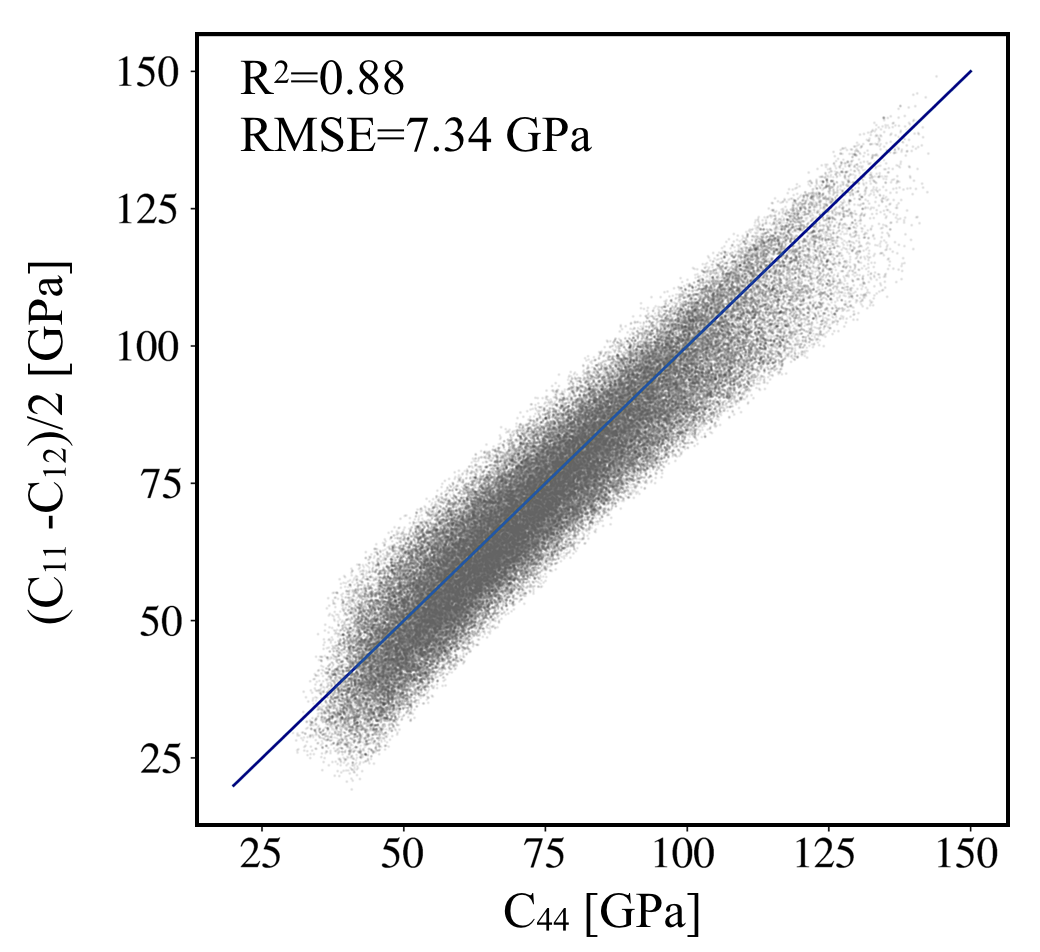}
    \caption{The  descriptor predicted C$_{44}$ was compared with C$_{44}$ estimated using independent elastic parameters, i.e., $\text{C}_{44}\approx(\text{C}_{11}-\text{C}_{12}$)/2, assuming zero anisotropy of disorder bcc Bravais lattice.}
\end{figure}

\subsection{High-throughput prediction of high strength HEAs and its physical origin.}

HEAs are expected to not only match the current state-of-the-art structural materials properties of conventional as well complex alloys, but they are expected to have lower density, optimal stiffness, i.e., low brittleness or higher ductility. However, the large composition space of HEAs restricts the exploration of physical space beyond equiatomic compositions. We show in Fig.~{7} that our computationally inexpensive analytical ML models facilitates quick search of technologically useful alloys over multi-dimensional spaces. 

{\par}The SISSO predicted analytical model of stiffness matrix (C$_{ij}$) (Eq.~{1} and see supplemental Table~1) were used to estimate the zero-temperature flow stress ($\tau_{y0}$), or critically resolved shear stress for edge dislocations in a BCC matrix, using the reduced-order model developed by Maresca and Curtin \cite{maresca2020mechanistic}. The main parameters ruling the strength model are the elastic constants and the misfit volume, the latter is calculated through Vegard's `law' using experimental values of atomic volumes. Although, our DFT results for solid-solution and associated elemental volumes could be used. The critically resolved shear stress is related to the yield-strength of the material through the Taylor factor, and so, for the current work, we used calculated $\sigma_{y0}$ for the yield-strength at 0 K. {As well as the approximation for finite-temperature (1300$ ^\circ C $), finite strain-rate (0.001) which builds up on the 0K approximation.}Our estimates of $\sigma_{0y}$ in Fig.~{7}a suggests towards a region in (Nb-Ta)-(Mo-W)-V compositions space that shows higher strength, which is different from high mixing entropy ($\Delta{S_{mix}}$) region (marked by solid white dot) doesn't necessarily correlate with mechanical or electronic behavior. In contrast to the usual expectations, however, we found that the region of high yield-strength in Fig.~{7}a directly correlates with higher  thermodynamic stability ($\Delta{H_{mix}}$) as shown in Fig.~{7}b. The increasing trends (ML predicted)  in yield-strength from quaternary NbTaMoW (1.12 GPa) to quinary NbTaMoWV (2.00 GPa) matches with experiments \cite{Senkov2011,Senkov2011a}, where adding V was found to increase the strength of the alloy.

\begin{figure}
    \centering
    \includegraphics[width=0.9\textwidth]{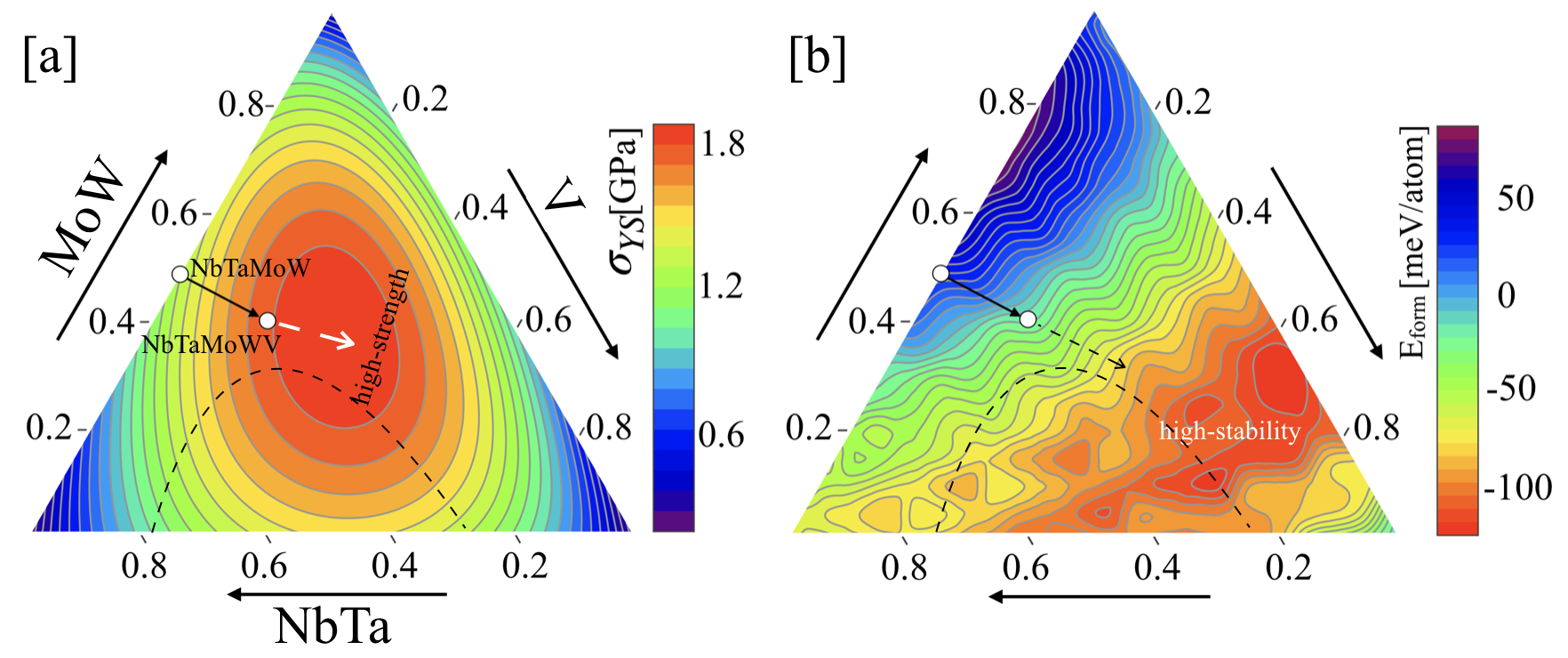}
    \caption{ (a) Yield-strength ($\sigma_{0y}$) and (b) formation enthalpy ($\Delta{H_{form}}$ at 0 K) vs  composition (c$_{i}$, i=Nb/Ta, Mo/W, V). The parabolic-dashed line shows the high electronegativity variance ($\Delta\chi_{var}$), while arrow-head from NbTaMoW to NbTaMoWV RHEA (solid-white circle) shows experimentally observed increasing trend. Our observation suggests that its not the single parameter but the interplay of different quantities, such as chemical entropy, $\Delta\chi_{var}$, and $\Delta{H_{form}}$, that are responsible for higher strength.}
\end{figure}

{\par}The dashed parabolic region marked in Fig.~{7}a,b shows the region of high electronegativity variance ($\Delta\chi_{var}$) (also see supplement Figure S1a). The $\Delta\chi_{var}$ increases with increase in V and Nb/Ta more compared to Mo/W because of higher  V electronegativity compared to others. We hypothesize that the large $\Delta\chi_{var}$ may lead to strong charge imbalance in the neighboring environment of V as large $\chi$ pulls more charge, which adds local lattice distortion and hence provide local solid-solution strengthening. The strong correlation between $\sigma_{0y}$ in Fig.~{7}a and $\Delta{H_{mix}}$ in Fig.~{7}b possibly arises from V addition that enhances the local-lattice distortion and the alloy stability,  as discussed by Song \etal~\cite{Song2017} and Singh \etal~\cite{Singh2021} Our assessment suggests that optimal combination of entropy (as a proxy for alloy complexity), phase stability, and  $\Delta\chi_{var}$ results into higher yield-strength.

{\par}The Young's moduli (E) is another important design characteristics that shows the tensile stiffness of materials such as HEAs. General convention suggests that maximizing E should also maximize the strength, e.g., high $\sigma_{0y}$. To understand this, we plot $E$  vs $\sigma_{0y}$  [in GPa] in Fig.~{8}a that shows a optimal E-region (150$<$E$<$250) in which $\sigma_{0y}$ is higher (marked by dashed lines), however, beyond the marked region, $\sigma_{0y}$ falls off quickly. We also found an inverse relation between E in Fig.~{8}b and $\sigma_{0y}$ in Fig.~{8}c with  $\Delta\chi_{var}$, i.e., E ($\sigma_{0y}$) decreases (increases) with increasing (decreasing)  $\Delta\chi_{var}$. This also confirms our idea of finding optimal tensile stiffness range rather than maximizing it while designing new RHEAs. 

\begin{figure}
    \centering
    \includegraphics[width=0.85\textwidth]{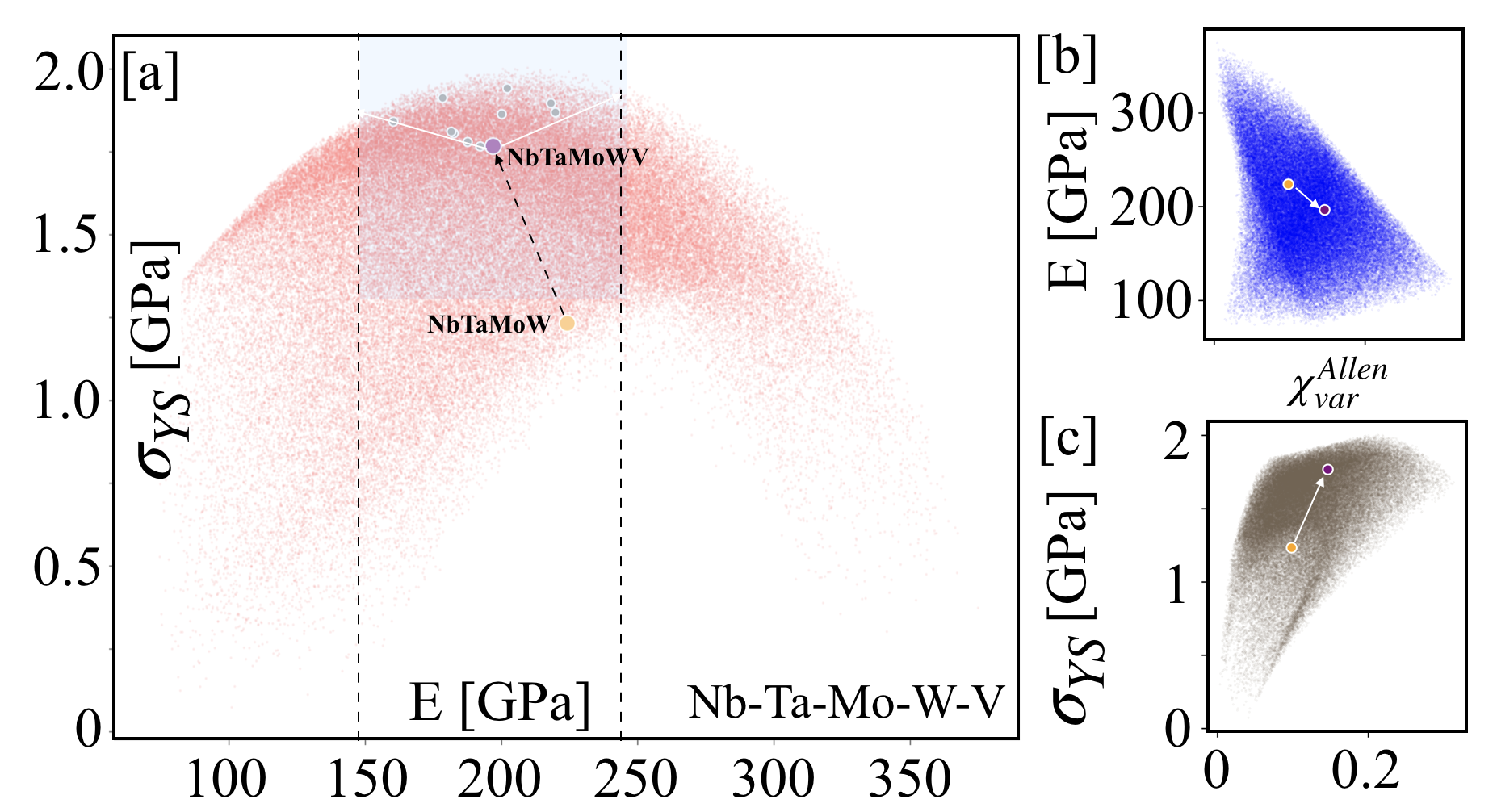}
    \caption{(a) Compositions with higher yield-strength ($\sigma_{0y}$; 1.5$<\sigma_{0y}<$2.25 GPa) away from high-entropy region (see Table~\ref{HEA_YS}) are highlighted vs optimal Young's moduli range (E; 150 $<$ E $<$ 250 GPa). (b,c) Electronegativity variance ($\chi_{var}$) was found to show strong correlation with (E, $\sigma_{0y}$).}
\end{figure}

{\par}From our analysis, we predict new RHEA compositions as shown in Table~\ref{HEA_YS}, where C$_{11}$, E, K, $\tau_{0y}$, and $\sigma_{YS}$ are tabulated along with  chemical composition of each alloying element. The RHEAs are arranged in order of  increasing density ($\rho$ in g/cc), which becomes critical depending on area of application. Notably, predicted RHEAs with higher $\sigma_{YS}$ directly correlates with higher V (20+ at.\%) concentration, this further emphasizes our hypothesis of optimal concentration of V in RHEAs improves both phase stability and mechanical behavior. This can be attributed to the fact that higher Allen-scale electronegativity of V pulls charges from neighboring sites and creates strong solid-solution strengthening through local lattice distortions. The size effect ($\delta$) plot in supplementary Figure 2b also shows the higher effect in the optimal region with higher $\sigma_{YS}$ marked in Fig.~{7}d. Moreover, alloys with higher V concentration in predicted RHEAs show increase in energy stability, see supplementary Table~2.

\begin{table}
	\begin{center}
		\caption{ML-predicted RHEAs with higher strength ($\sigma_{0y}$; GPa) compared to equiatomic MoNbTaW and MoNbTaVW. Results are arranged in increasing density.}
		\scalebox{0.75}{
			\begin{tabular}{c|ccccc|ccccccc}
				\hline 
				\multirow{1}{*} {MPEAs}  & Mo  & Nb  & Ta & V & W  & $\rho$ & C$_{11}$ & E & K &  $\sigma_{0y}$ & $\sigma_{1300C}$ &G/K\tabularnewline
				&   &   & [at.\%] &  &   & [g/cc] & [GPa] & [GPa]  & [GPa]  & [GPa] & [GPa]  & \tabularnewline
				\hline 
				MoNbTaW   & 25  & 25  & 25  & $-$       & 25  & 13.68  & 344  & 224 & 235  & 1.23 &0.22&0.33\tabularnewline
				MoNbTaVW   & 20  & 20  & 20  & 20 & 20  & 12.17 & 320  & 197 & 223 & 1.76 & 0.35&0.31\tabularnewline	
				\hline
				RHEA0  & 19.99  &   0.70  & 30.02  & 44.91  &   4.38  & 10.69 &285	&160	&205&	1.842 &0.29&0.268\tabularnewline
				RHEA1   & 19.57  &   8.83  & 13.76  & 40.82  & 17.02  & 10.83 &311&	182	&218&	1.807 &0.32&0.295\tabularnewline
				RHEA2   & 17.68  &  9.08   & 10.84  & 41.31  & 21.09  & 10.99 & 317& 187& 220 	&1.781 &0.32&0.302\tabularnewline
				RHEA3   &   4.67  & 15.32  &   6.17  & 42.79  & 31.05  & 11.43 & 312&	182&	220	&	1.811 &0.32&0.293\tabularnewline
				RHEA4   &   8.64  &   1.94  & 14.54  & 46.28 &  28.60  & 11.82 & 318 & 192 & 223	&1.767 &0.31&0.302\tabularnewline
				RHEA5   & 16.84  &  2.09   & 34.60  & 34.75  & 11.72  & 12.05 & 298	& 178 & 213&	1.913 & 0.35&0.285\tabularnewline
				RHEA6   &   3.17  & 23.98  &   5.89  & 28.22  & 38.74  & 12.56 & 328	&200&	229	&1.864 & 0.38&0.317\tabularnewline
				RHEA7   & 10.31  &  8.63   & 12.87  & 31.20  & 36.99  & 12.98 & 343 & 219	& 235 & 1.870 & 0.40&0.334\tabularnewline
				RHEA8   & 23.31  &  1.31   & 32.20  & 22.91  & 20.27  & 13.17 & 335	& 218&230	&1.897 & 0.41&0.324\tabularnewline
				RHEA9   &   1.07  &  0.78   & 40.59  & 26.80  & 30.76  & 14.51 & 313&202&226&1.942 &  0.41&0.300\tabularnewline
				\hline 
		\end{tabular}}
		\label{HEA_YS} 
	\end{center}
\end{table}

{\it Discussion on ductility:~} The criteria to evaluate the ductile behavior based on elastic constants in BCC HEA has been reported at several instances.\cite{tian2014ab,Pugh1954,gu2006critical} However, it is impossible that all material properties or all type of alloys can be fit using same kind of model or approach. Nonetheless, the Pugh ratio offers good correlation with in ductility of materials \cite{Pugh1954}, and, as noted above, supported by  existing experimental data from large number of materials \cite{gschneidner2003family}. Indeed, this is reflected in our data, see  Figs.~8a and ~9a. Poisson's ratio $\nu$, considering the relationship between $\nu$ and (G/K), has the potential to assess the ductile behavior of MPEAs (Fig.~9b), where the brittle-to-ductile transition limit  corresponds to $\nu > 0.26$ \cite{lee2020temperature}. Similarly, the large positive Cauchy pressure in Fig.~9c represents an increased degree of the metallic bonding that correlates with ductility \cite{lee2020temperature}, whereas the negative value is indicative of  brittleness (or more covalent bonding).\cite{pettifor1992theoretical} Recently, Lee \etal~\cite{lee2020temperature} have reported ductility in NbTaTiV RHEA, which further supports our idea that state-of-the-art, ab-initio calculations and machine-learning approaches may be helpful for refractory-based alloys. This also establishes the quality of prediction of high-strength regions in Fig.~7a.

\begin{figure}
    \centering
    \includegraphics[width=0.8\columnwidth]{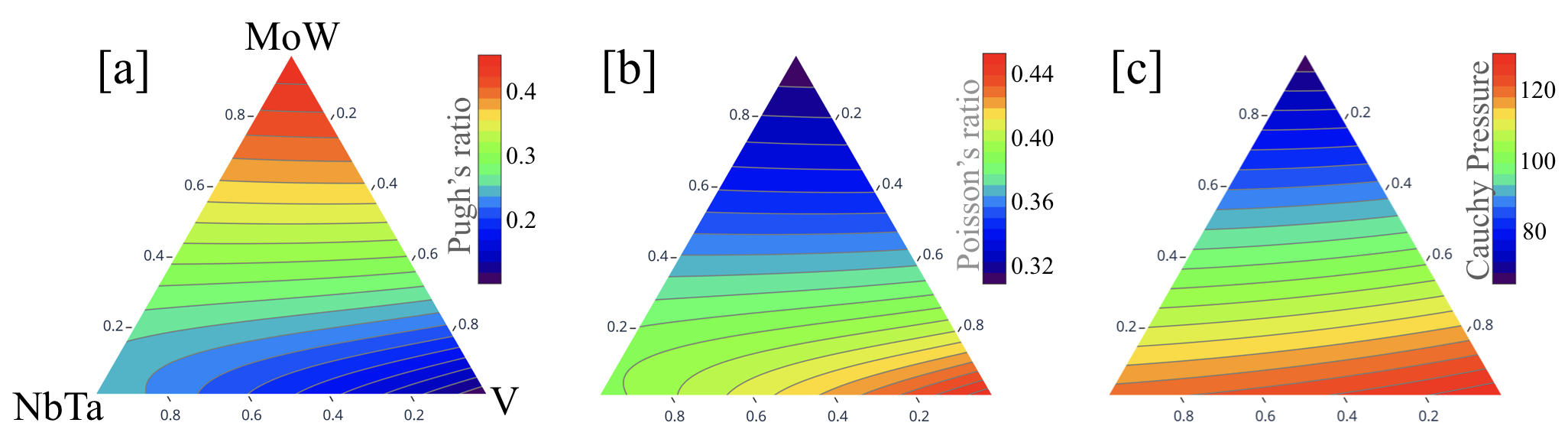}
    \caption{We plot the (a) Pugh's ratio, (b) Poisson's ratio, and (c) Cauchy pressure (C$_{12}$-C$_{44}$ in GPa) that highlights  the optimal region of strength lies slightly away from high-entropy region.}  
\label{fig:sisso_1}
\end{figure}

\subsection{Application of the proposed SISSO model to systems outside the training chemistry.}

In Table ~\ref{tab_ms}, we demonstrate the application of the SISSO models to several stoichiometric HEA alloys with chemistries outside the chemical space used to train the model in the first place. {As shown in Table ~\ref{tab_ms}, we found good agreement between descriptor predicted elastic constants and DFT. If we focus on ML vs DFT results in Table~2, error in prediction for most cases is within 1-5\%. Notably, the C$_{ij}$ for NbTaTiVZr is the only exception that shows large error, i.e., C$_{11}$ (171 vs 190 GPa, i.e., $\sim$10\% error), C$_{12}$ (150 vs 104 GPa, i.e., $\sim$44\% error), C$_{44}$ (10 vs 43 GPa, i.e., $\sim$77\% error). In our limited understanding to this aspect, we believe the possible reason in under predicting C$_{ij}$'s can be two-fold $-$ (i) Ti/Zr not included in the training datasets, and (ii) crystal anisotropy of Ti/Zr (hexagonal parent phase). Here, we mainly attribute this fact to crystal anisotropy of Ti/Zr as C$_{ij}$ are strongly direction dependent, which is not included in our training dataset. Despite minor disagreements, these results are promising, as they tentatively suggest that the model is capable of making safe extrapolations.} We want to emphasize that more insights are needed into the extrapolation of the SISSO based descriptors and the chosen featurization for the complete generalization of the proposed descriptor. A more thorough investigation, beyond the scope of this work, is necessary to ascertain the ability of this model to extrapolate, with reasonable accuracy, beyond the training chemistry.

\begin{table}
	\caption{The C$_{11}$, C$_{12}$, C$_{44}$, bulk-moduli, and elastic-moduli of diverse HEAs (some outside the training chemistry) predicted from the SISSO model show good agreement with DFT.}
	\label{tab_ms}
	\begin{tabular}{llrlrrlrlrl}
		\hline
		System & \multicolumn{2}{l}{$C_{11}$} & \multicolumn{2}{l}{$C_{12}$} & \multicolumn{2}{l}{$C_{44}$} & \multicolumn{2}{l}{K}        & \multicolumn{2}{l}{E}        \\ \hline
		& ML & \multicolumn{1}{l}{DFT} & ML & \multicolumn{1}{l}{DFT} & \multicolumn{1}{l}{ML} & DFT & \multicolumn{1}{l}{ML} & DFT & \multicolumn{1}{l}{ML} & DFT \\
		MoNbTaVW  & 321 & 328 & 172 & 169 & 74  & 79  & 223 & 222 & 197 & 213 \\
		CrMoNbTaV & 315 & 330 & 160 & 159 & 77  & 85  & 213 & 216 & 206 & 226 \\
		CrMoNbTaW & 370 & 378 & 172 & 168 & 99  & 105 & 241 & 238 & 244 & 275 \\
		CrMoNbVW  & 363 & 372 & 169 & 170 & 97  & 101 & 236 & 237 & 235 & 265 \\
		CrMoTaVW  & 372 & 393 & 171 & 170 & 100 & 111 & 242 & 244 & 248 & 290 \\
		CrNbTaVW  & 322 & 360 & 166 & 156 & 78  & 102 & 222 & 224 & 204 & 266 \\
		NbTaTiVZr & 171 & 190 & 150 & 104 & 10  & 43  & 146 & 132 & 116 & 116  \\ \hline
\end{tabular}
\end{table}

\subsection{Analytical model for misfit-volume and yield-strength}

While Maresca and Curtin \cite{maresca2020mechanistic} estimated the misfit volume using experimental values, all the numbers and details are not clearly stated. We emphasize that the misfit volume calculated with different atomic volumes are expected to show deviations when used with the reduced strength model, as shown in Table~3. Another source of uncertainty comes from elastic constants determined from the rule of mixing and the SISSO descriptors. However, descriptors-based analytical models are more close to reality as all possible alloying effects are included, while the rule of mixtures are not.

\begin{table} 
\caption{The misfit-volume and strength calculation with different elemental misfits results into expected deviations.}
\resizebox{\columnwidth}{!}{%
\begin{tabular}{|lllllllllll|}
\hline
\multicolumn{1}{|l|}{} &
  \multicolumn{1}{l|}{$\Delta V_{Mo}$} &
  \multicolumn{1}{l|}{$\Delta V_{Nb}$} &
  \multicolumn{1}{l|}{$\Delta V_{Ta}$} &
  \multicolumn{1}{l|}{$\Delta V_{V}$} &
  \multicolumn{1}{l|}{$\Delta V_{W}$} &
  \multicolumn{1}{l|}{$\sum_{n} c_{n} V_{n}^{2}$} &
  \multicolumn{1}{l|}{$\tau_{y,0}$ (GPa)} &
  \multicolumn{1}{l|}{C$_{11}$} &
  \multicolumn{1}{l|}{C$_{12}$} &
  C$_{44}$ \\ \hline
\multicolumn{11}{|l|}{Mo-Nb-Ta-V-W} \\ \hline
\multicolumn{1}{|l|}{Maresca and Curtin's reduced model} &
  \multicolumn{1}{l|}{-0.628} &
  \multicolumn{1}{l|}{1.713} &
  \multicolumn{1}{l|}{1.877} &
  \multicolumn{1}{l|}{-2.484} &
  \multicolumn{1}{l|}{-0.478} &
  \multicolumn{1}{l|}{2.650} &
  \multicolumn{1}{l|}{0.610} &
  \multicolumn{1}{l|}{346.8} &
  \multicolumn{1}{l|}{157.7} &
  90.5 \\ \hline
\multicolumn{1}{|l|}{Vegard's law (misfit factor)/SISSO (elastic constants)} &
  \multicolumn{1}{l|}{-0.681} &
  \multicolumn{1}{l|}{1.760} &
  \multicolumn{1}{l|}{1.810} &
  \multicolumn{1}{l|}{-2.424} &
  \multicolumn{1}{l|}{-0.465} &
  \multicolumn{1}{l|}{2.586} &
  \multicolumn{1}{l|}{0.576} &
  \multicolumn{1}{l|}{320.8} &
  \multicolumn{1}{l|}{172.0} &
  73.0 \\ \hline
\multicolumn{11}{|l|}{Mo-Nb-Ta-W} \\ \hline
\multicolumn{1}{|l|}{Maresca and Curtin's reduced model} &
  \multicolumn{1}{l|}{-1.293} &
  \multicolumn{1}{l|}{1.135} &
  \multicolumn{1}{l|}{1.168} &
  \multicolumn{1}{l|}{-} &
  \multicolumn{1}{l|}{-1.010} &
  \multicolumn{1}{l|}{1.336} &
  \multicolumn{1}{l|}{0.450} &
  \multicolumn{1}{l|}{375.5} &
  \multicolumn{1}{l|}{167.3} &
  101.6 \\ \hline
\multicolumn{1}{|l|}{Vegard's law (misfit factor)/SISSO (elastic constants)} &
  \multicolumn{1}{l|}{-1.287} &
  \multicolumn{1}{l|}{1.154} &
  \multicolumn{1}{l|}{1.204} &
  \multicolumn{1}{l|}{-} &
  \multicolumn{1}{l|}{-1.071} &
  \multicolumn{1}{l|}{1.396} &
  \multicolumn{1}{l|}{0.402} &
  \multicolumn{1}{l|}{344.2} &
  \multicolumn{1}{l|}{182.0} &
  86.2 \\ \hline
\end{tabular}
}
\end{table}

We developed the analytical SISSO model to predict alloy misfit volume factor (see method section) trained on same binary and ternary, which is similar to analytical descriptors in Eq.~1-3. The misfit volume descriptor can directly be used with  Maresca and Curtin's reduced strength model.\cite{maresca2020mechanistic} In Fig.~10a, the 10-fold cross-validation test shows the accuracy of 0.89 (R$^{2}$) with an RMSE of 0.38. While this is a good value for predictions within the training data, the uncertainty transmitted to the strength model may grow larger. The model presented in Eq.~4 shows that accurate analytical models for misfit volumes can be developed, however, we need more accurate high-fidelity data from experiments or ab-initio methods.

\begin{gather} 
\begin{aligned}\label{eq:young}
\scalebox{0.9}{$ V^{misfit} = -0.0654-93800\frac{a_{diff}}
{\chi_{Pauling}^{avg}\sqrt{a_{avg}}} +376000\frac{a_{diff}}{\chi_{Pauling}^{avg}
\left( a_{avg}+a_{red}\right)}  -132000 \frac{a_diff}{\chi_{Pauling}^{avg} a_{avg} a_{red}}$}
\end{aligned}
\end{gather}

\begin{figure}
    \centering
    \includegraphics[width=0.75\textwidth]{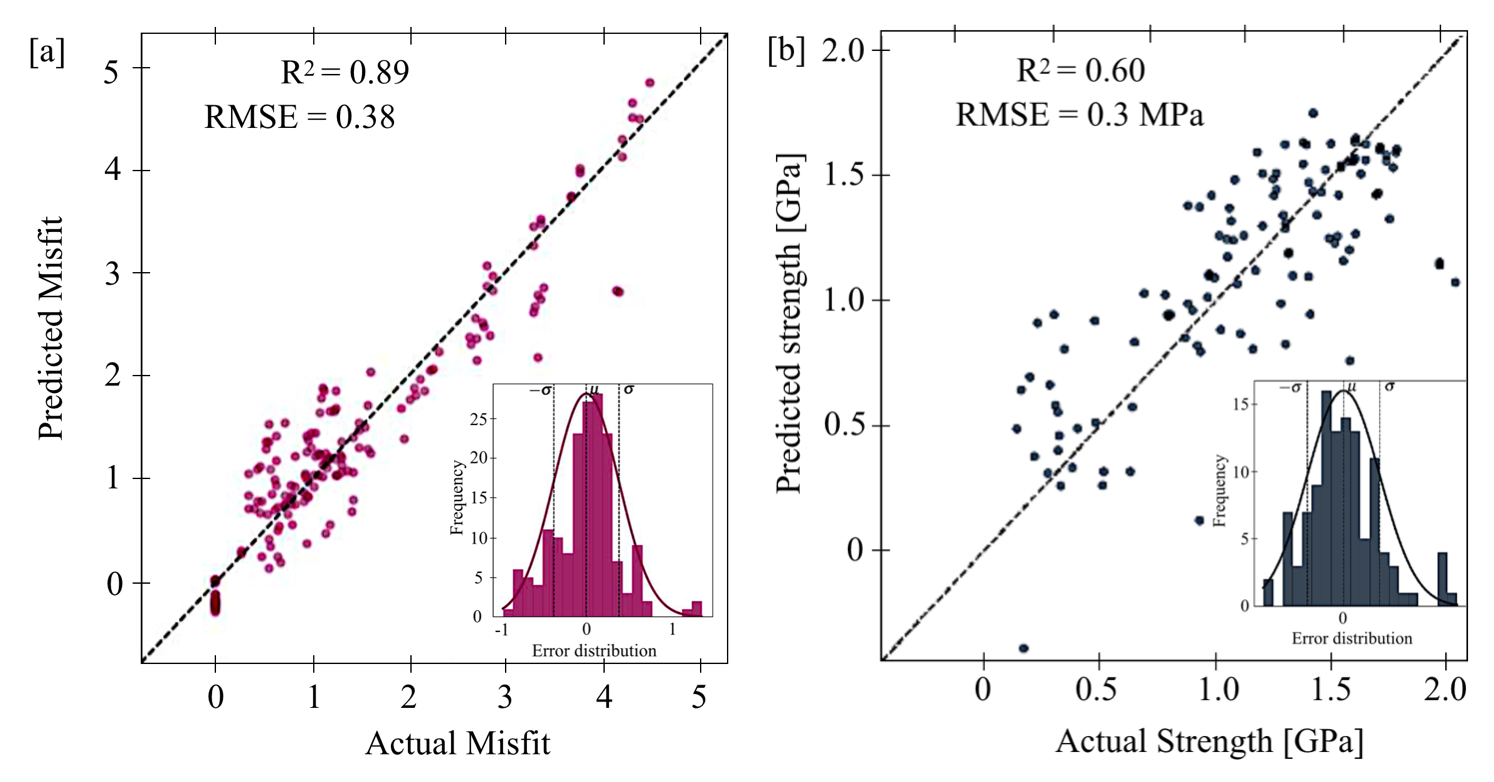}
    \caption{Actual vs predicted (a) misfit volume, and (b) strength ($\sigma$; GPa) from ten-fold cross-validation test. Inset in a,b shows the error distribution around the mean, which is virtually zero for all the three cases.}
\end{figure}

We  further emphasize that the ML-trained strength models for MPEAs are not so readily available in literature due to the lack of the training dataset. However, we could develop a sparse strength database for HEAs consisting of 114 experimental HEAs (fcc, bcc, and fcc+bcc). The featurization used was also alloy specific but simpler, using only averages and thermodynamical parameters. A 10-fold cross-validation test was used to showcase the overall confidence of the SISSO model in Fig.~10b. An error using the 3D descriptor is prone to overfitting as cross-validation accuracy dropped to 0.60 from 0.74 (trained on all whole dataset). We believe that the possibly of large comes from limited experimental data with unknown experimental conditions, sample type and quality, and other unknown critical quantities. This indicates towards the need of more reliable dataset creation from well-defined experimental sources and conditions for robust machine-learning strength model. This discussion highlights the need of intelligent design in materials discovery, i.e., in absence of reliable data mean-filed approaches can be used for accelerated  filtering of multi-dimensional hyperspace such as high-entropy alloys.

{\par} {\it Caveats in DFT vs ML-predicted shear:}~ In refractory-based alloys, the shear or trigonal elastic parameter (C$_{44}$) of individual elements, especially, Nb is underestimated in most electronic-structure methods (DFT: 17 GPa \cite{Ahuja2008}; Expt: 31 GPa \cite{Eriksson1993}),  independent of DFT methods or exchange-correlation functional \cite{Nagasako2010}. This issue has been discussed a length in literature, {\it e.g.}, by  Koci \etal~\cite{Ahuja2008}, who attribute it to transverse-phonon mode-driven anomalous band dispersion; more recently, as DFT typically freezes core states, it was associated with warping of the low-lying core levels during shearing, which was not addressed previously. Nonetheless, this issue limits the reliable data generation for RHEAs. And, interestingly, our SISSO-trained analytical model shows better accuracy for Nb, i.e., 24 GPa compared to experiments (31 GPa) \cite{Eriksson1993}.

\section{Summary and discussion}

{\par} In this work, descriptor-based machine-learning (ML) framework models were developed to efficiently scan and predict HEA elastic properties. The  power of descriptor-based analytical models for fast exploration of the was exemplified for refractory based Nb-Ta-Mo-W-V  HEAs. The reliable, optimal, and interpretable analytical descriptors were trained using SISSO method using a database of elastic properties yielded from density-functional theory calculations.  A detailed analysis of target properties was also carried out to correlate  common elemental/alloy features for optimized descriptor  to better interpret proposed analytical models, distinctly different from black-box ML models. 

{\par}The descriptor-predicted stiffness matrix (C$_{ij}$) of Nb-Ta-Mo-W-V HEAs were used to assess technologically useful quantities, such as yield-strength, that identify high-strength regions that correlate more with optimal combination of entropy (not high-entropy region), regions with large size effect ($\delta$), large electronegativity variance ($\chi_{var}$), and  regions of high phase stability (lower formation energy). Our predicted trends match limited existing experiments, further establishing inexpensive descriptor-based methods can accelerate design of technologically useful alloys.  

{\par}Elastic relations, like trends in Cauchy stability (C$_{12}$-C$_{44} <0$), show similar behavior as yield-strength in the quinary space suggesting a more direct approach to estimate ductility using analytical descriptors. To further emphasize this point, we note that the alloy compositions with excellent mechanical properties is not necessarily those with highest chemical entropy or valence-electron count. For example, compositions with high-strength and high phase-stability (lower formation enthalpy) were found in lower-entropy regions, again questioning the focus on maximizing entropy to achieve better mechanical behavior.

{\par}Our results emphasize that computationally inexpensive models are important for thorough and accurate search of the vast HEA composition space to identify regions with desirable target properties. For that, an unconstrained search of the alloy space permits a further possibility to optimize HEAs compositions. The application of the model to chemistries (beyond those used to train it) are promising, although additional work is needed to ensure the safety of such extrapolations. Regardless, the model is considered to be useful for further design in the Nb-Ta-Mo-W-V system.

\section*{Acknowledgements}
ARPA-E ULTIMATE project \emph{Batch-wise Improvement in Reduced Design Space using a Holistic Optimization Technique} (BIRDSHOT) under primary contract No. DE-AR0001427 is acknowledged. GV, KY and RA acknowledge the support of QNRF under Project No. NPRP11S-1203-170056. RC was supported by NSF-CMMI Grant No. 1663130. DS was supported by NSF Grant No. 1545403. Methods developed at Ames Laboratory (DDJ and PS) and used herein were supported by the U.S. Department of Energy (DOE), Office of Science, Basic Energy Sciences, Materials Science \& Engineering Division. Ames Laboratory operated by Iowa State University for the U.S. DOE under contract DE-AC02-07CH11358. Calculations were conducted using the advanced computing resources provided by Texas A\&M High-Performance Research Computing.

\section*{Competing Interests}
The authors declare no competing interests.

\section*{Data availability}
The authors declare that the data supporting the findings of this study are available within the paper and supplement. Also, the data that support the plots within this paper and other findings of this study are available from the corresponding authors upon reasonable request. Supporting data and code for all figures generated using descriptors in Eq. 1-3 are available at {\sl Github}~\cite{elasticsisso}.

\section*{Correspondence}
Correspondence should be addressed to GV (guillermo.vazquez@tamu.edu), PS (psingh84@ameslab.gov) or RA (raymundo.arroyave@tamu.edu).

\section*{References}
\bibliography{Manuscript_SISSO}
\end{document}